\documentclass[11pt,a4paper]{article}

\usepackage[utf8]{inputenc}
\usepackage[T1]{fontenc}
\usepackage{lmodern}

\usepackage{amsmath}
\usepackage{amssymb}
\usepackage{amsthm}

\usepackage{graphicx}
\usepackage{booktabs}
\usepackage{multirow}
\usepackage{array}

\usepackage{algorithm}
\usepackage{algorithmic}
\usepackage{float}

\usepackage{xcolor}

\usepackage{hyperref}
\hypersetup{
    colorlinks=true,
    linkcolor=blue,
    citecolor=blue,
    urlcolor=blue
}

\usepackage{orcidlink}

\usepackage[margin=1in]{geometry}

\newtheorem{definition}{Definition}
\providecommand{\keywords}[1]{\par\medskip\noindent\textbf{Keywords:} #1}

\newcommand{\R}{R}
\newcommand{\Srel}{S}
\newcommand{\Ji}{J_{ij}}
\newcommand{\M}{\mathcal{M}}
\newcommand{\Ahat}[1]{\hat{A}_{#1}}

\title{Join Indices for Search Engines:\\ a Prunable Parallel Semijoin over Lucene Segments}

\author{
    Mikhail Khludnev\orcidlink{0009-0007-2738-0153} \\
    Reksoft\\
    \texttt{mkhl@apache.org}
}

\date{August 2026}

\begin{document}
\maketitle

\begin{abstract}
Joins are second-class citizens in search engines: existing query-time
join implementations in Lucene are limited either in performance or in
capability, forcing a choice between fast joins scoped to a single index
and slower joins that span independently managed indices. We carry
Valduriez's join-index technique from relational systems to Lucene's
flush-based (LSM-style) segment storage: for every pair of a parent and a
child segment we materialize an append-only, ordinal-to-ordinal
join-index column $\Ji[c]=p$, avoiding any query-time translation of
external variable-length keys. On top of this structure we build a
semijoin algorithm that is computed per parent segment, in parallel,
without a global barrier between stages; it prunes at three levels
(segment-level, the first of which comes free from per-segment execution;
a-priori min/max; and document-level two-phase confirmation with a lazily
accumulated \emph{half-read union}) so that it
composes with arbitrary engine queries instead of wasting computation on
matches that a sibling filter would later discard. A prototype
implemented as an Apache Solr query parser, benchmarked on 1M products
joined against 10M skus, cuts average query latency $5.4\times$
(359.8\,ms vs.\ 1934.6\,ms) relative to Solr's built-in query-time join, and
the advantage widens monotonically with load, reaching $8.3\times$ at a
concurrency of eight: on 4 vCPUs the baseline peaks at 1.18 queries/s and then
loses throughput, while the join index is still gaining, at 8.04 --- $6.8\times$
the baseline's best. Both arms return identical result counts on every query of
every run. Instrumenting that run places the credit on the structure rather than
on the pruning built over it: the benchmark draws foreign keys uniformly at
random, which flattens every column's bounding ranges to the full parent
segment, and the two range-based levels consequently fire on under $2\%$ of
segment pairs. The speedup is thus the join index's own, measured on a workload
that defeats its own optimizations. What we establish for the pruning hierarchy
is composability; what it delivers on correlated key distributions remains to be
measured.
\end{abstract}


\section{Introduction}
\label{sec:intro}

Inverted indexes~\cite{manning2008iir} are the standard technology behind
full-text and faceted search --- powering web search engines, digital
library catalogs, and e-commerce product search alike. Their key
feature is the picking of the top $k$ relevant results. Many
optimizations known as \emph{dynamic pruning} have been applied to this algorithm.
We describe them closely in \S\ref{sec:dynamicpruning} using Apache Lucene\footnote{Apache Lucene and Lucene are trademarks of the Apache Software Foundation.}~\cite{lucene} as
an example implementation.

Relational databases, on the other hand, provide highly performant
join algorithms. Join operations in search engines remain a less
developed area, and existing implementations are limited in either
performance or capability, or both.

The task of this paper is therefore:
\begin{itemize}
    \item to implement a fast join algorithm for inverted index, but only assuming equijoin in a form of semijoin~\cite{bernstein1981semijoins} (see the SQL sketch in \S\ref{sec:problem});
    \item to make it compatible with dynamic pruning algorithms of search engines;
    \item to make it suitable for parallel execution to achieve high performance and scalability.
\end{itemize}

This paper uses Apache Lucene~\cite{bialecki2012lucene4} as a concrete example of the search engine,
but the ideas are applicable for other search engines and many NoSQL databases, which share key design elements.

\section{Contributions}
\begin{itemize}
    \item applying Valduriez's join-index~\cite{valduriez1987} idea to 
    inverted index, meaning the flush-produced, unmerged runs of an LSM-tree~\cite{oneil1996lsm} or L0 in terms of RocksDB~\cite{dong2017rocksdb}, using Apache Lucene as a concrete example;
    \item the semijoin algorithm suitable for parallel execution, with no global
    barrier between stages;
    \item a three-level pruning hierarchy (segment, a-priori, document) --- the
    first of which comes free from per-segment execution --- which lets the
    semijoin enter an existing dynamic-pruning engine as an ordinary conjunct.
    The claim here is composability, not effectiveness:
    \S\ref{sec:pruninginstrumentation} measures the two range-based levels
    firing on under $2\%$ of segment pairs on our benchmark, and what they are
    worth on correlated key distributions is left open
    (\S\ref{sec:futureeval});
    \item an instrumented evaluation that separates those two questions, and
    attributes the measured speedup to the join-index column and the
    per-segment execution rather than to the pruning built on top of them
    (\S\ref{sec:pruninginstrumentation}).
\end{itemize}

\section{Results Summary}
Section~\ref{sec:experiments} benchmarks a prototype implementation, built as an
Apache Solr\footnote{Apache Solr and Solr are trademarks of the Apache Software Foundation.} query parser, against Solr's existing cross-collection join,
\texttt{\{!join score=none\}}. On an index of 1M products joined against 10M
skus, over 500 queries per arm, the join-index approach cuts average query
latency $5.4\times$ at a concurrency of one (359.8\,ms vs.\ 1934.6\,ms), rising
to $8.3\times$ at a concurrency of eight. Swept across concurrencies 1 to 8 on
4 vCPUs, the baseline peaks at 1.18 queries/s and then declines, while the join
index is still gaining at 8.04 --- $6.8\times$ the baseline's best figure at any
concurrency, and $7.2\times$ its figure at the same one. Both arms execute the
same query set and agree on the result count of every query.

Section~\ref{sec:pruninginstrumentation} then instruments that same run to ask
which part of the design earns the speedup, and the answer is not the part we
set out to claim. The benchmark's foreign keys are drawn uniformly at random,
which flattens every join-index column's bounding ranges to the full parent
segment; the a-priori level consequently bypassed $1.6\%$ of segment pairs and
the document level spared $1.5\%$ of column reads. The speedup is therefore the
join index's own --- the ordinal-to-ordinal column that removes query-time key
translation, and the per-segment, barrier-free execution --- obtained on a
workload that happens to defeat the pruning layered over it. We report the
pruning hierarchy accordingly: as a composability property, always available and
cheap enough that it cost nothing visible here, whose \emph{effectiveness} under
correlated key distributions is a question \S\ref{sec:futureeval} is designed to
answer and this paper does not.

\section{Background: the Execution Model}
\label{sec:background}

\subsection{Flush-produced segments}
\label{subsec:segments}
An Apache Lucene index~\cite{bialecki2012lucene4} consists of segments (or extents in common terms) and
might be viewed as an LSM-tree~\cite{oneil1996lsm,liang2026writeread}, but specifically it is a flush-produced
unmerged runs. These segments are autonomous inverted indices and suitable for independent
parallel searching that is already done in
Apache Lucene and Lucene-based engines such as
OpenSearch\footnote{OpenSearch is a registered trademark of Amazon Web Services, Inc.}~\cite{opensearch_concurrent_segment_search}, Apache Solr, and others. Note: a
\emph{slice} --- a unit of concurrent processing
--- might be less or more than a segment, but this detail is transparent to this paper.

\subsection{Virtual IDs}

Lucene is a system with virtual IDs~\cite{abadi2013columnstores}.
Adopting a notation akin to relational algebra, an index $\R$
consists of $N$ disjoint segments, $\R = R_1 \uplus R_2 \uplus
\dots \uplus R_N$; within segment $R_i$, a document is identified
by an ordinal (virtual docid) document number $d \in [0,
\mathit{maxDoc}_i)$, borrowing Lucene's own
\texttt{maxDoc()}~\cite{lucene_segments_docids} terminology ---
written $|R_i|$ in the set-based notation of
Section~\ref{sec:problem}, which extends this scheme to the child
relation $\Srel$. 

In the remainder of this paper, we use the term \emph{docid} to refer to these segment-scoped virtual IDs.

Note that docids differ from physical record offsets: 
\emph{surrogates}~\cite{valduriez1987} and RID~\cite{oneil1997} share 
a property of immutability and internal assignment, but might be processed 
efficiently as ordinals.

For now, we set aside the deletion marks as not significant to the core ideas.

There are two consequences of using virtual IDs (docids):
\begin{itemize}
    \item Since docids are assigned sequentially, they are dense
          and narrow, and compress better than the wide, sparse
          external IDs~\cite{abadi2013columnstores}.
          Compression is, in fact, a cornerstone of Lucene's
          performance~\cite{bialecki2012lucene4}. 
    \item Since segments are immutable once written (deletions
          aside), segment-scoped docids remain stable for the
          segment's lifetime.
\end{itemize}

\subsection{Value Ordinals and Dictionary Compression}
\label{sec:valueordinals}

Dictionary compression is a technique commonly used in columnar
stores~\cite{abadi2013columnstores}. When storing a value by docid, \texttt{values[docid]}, we
encounter two issues: variable-length values are harder to look
up, and repeating values waste storage and impact processing
time. Instead, we assign dense, narrow, fixed-width ordinals to
values and map each docid to its ordinal, hence obtaining a value by docid becomes
\texttt{values[ordinals[docid]]}. We can say that dictionary
compression introduces virtual IDs for values. Also, see a note in \S~\ref{app:invidx_dict}.

However, when there are many segments, 
the same values are assigned different ordinals in different segments. 
Thus, value ordinals can't be compared across segments, 
which causes a challenge for join implementation, 
since joins work on value equality. 
That can be addressed with two distinct measures:
\begin{itemize}
    \item avoid dictionary compression by limiting users to
    fixed-width numerics, which can simplify query-time join
    (e.g.\ using integers for FK and PK) --- but we consider it
    too restrictive and do not adopt this workaround in this paper;
    \item restrict join sides to a single index and single field and 
    correlate them via global ordinals (see \S\ref{subsec:otherjoins}) ---
    again, we put it aside as too restrictive for this paper.
\end{itemize}

\subsection{Dynamic pruning}
\label{sec:dynamicpruning}

There are many optimizations applied to top-$k$ retrieval, such as:
\begin{itemize}
\item adaptive (``leapfrog''/galloping) intersection for conjunctive
queries~\cite{demaine2000}, which skips runs of non-matching entries
between two sorted posting lists by exploiting the gaps between their
current positions, rather than merging them element by element;
\item the two-phase iteration algorithm~\cite{lucene6198}, where a
cheap approximation first narrows the candidate set with \emph{false-positives}, and the
expensive exact match --- e.g.\ confirming a phrase's term positions
--- is evaluated only on candidate documents that satisfy the cheaper
checks first;
\item MaxScore~\cite{turtle1995}, later refined into block-max
variants such as Block-Max WAND~\cite{ding2011bmw}, which use a
per-block upper bound on relevance to skip entire runs of documents
that provably cannot outscore the current threshold, as implemented
in Apache Lucene and borrowed by search servers based on it, such as Elasticsearch\footnote{Elasticsearch is a trademark of Elasticsearch B.V., registered in the U.S. and in other countries.}~\cite{elastic_bmw}, Apache Solr, etc.
\end{itemize}

Our task is an efficient semijoin: an algorithm shaped as a segmented
iterator over docids that takes another segmented iterator as
input. This is what a semijoin implementation needs. For dynamic
pruning, we design this algorithm to compose with arbitrary engine
queries and their optimizations.

\section{Related Work and Design Space}
\label{sec:related}

\subsection{The classic query-time join}
\label{sec:qtimejoin}

The existing implementation of query-time
join~\cite{lucene_query_time_joins} in Apache Lucene is, in relational
terms, a block index join~\cite{graefe1993query}. We do not use that name
in the remainder of the paper: \emph{block} would collide with Lucene's own
Block Join (\S\ref{subsec:otherjoins}), which is an entirely different
algorithm. We refer to it throughout as the \emph{classic query-time join},
always with a pointer to this section.
Consider the following sample semijoin query:

\texttt{SELECT * FROM R WHERE EXISTS (SELECT 1 FROM S
WHERE S.fk = R.id)}

It is executed in two \emph{stages}:
\begin{enumerate}
    \item loop through child ($\Srel$) records, collecting foreign
          keys (FKs) into a set;
    \item bulk-search that FK set against the primary key field
          \texttt{R.id}.
\end{enumerate}
We say \emph{stages} rather than \emph{phases} deliberately: \emph{phase} is
reserved in this paper for Lucene's two-phase iteration
(\S\ref{sec:dynamicpruning}, \S\ref{subsec:columnsandquery}), the
approximation/confirmation contract that our own algorithm implements in
\S\ref{sec:twophase}. The two notions are unrelated, and the collision is
easy to make.

Given that the indexes are partitioned across segments
(\S\ref{subsec:segments}), both stages can exploit per-segment
parallelism, but in the middle we need to merge the set of
$\{FK\}$ sequentially. That limits the scalability according to
Amdahl's law~\cite{amdahl1967validity}.
The motivation for this two-stage design is that Lucene has an optimized bulk search operation; avoiding the
sequential merge would lead to multiple sub-search operations
that are expected to be slow --- but this has not been evaluated.

\subsection{Other join algorithms in Apache Lucene}
\label{subsec:otherjoins}
We mention them briefly for completeness, although they are not directly related to our work:
\begin{itemize}
\item Index-time join~\cite{harwood_nested_documents} --- a rather fast algorithm,
since parents and their children are indexed together as a contiguous block of
adjacent docids. That physical clustering is exactly what makes it prone to
update amplification: touching a single child forces the whole block to be
reindexed, which rules the approach out for many real-life scenarios we want to
pursue. Its closest analog we can come up with is the hierarchical storage of IBM
IMS~\cite{ibm_ims_sequential} --- a design that goes back to 1968, predating the
relational model~\cite{codd1970} itself, and is still shipping today --- where dependent segments
are stored beneath their parent and traversed in hierarchical sequential
order. It is also unfortunate
that Lucene calls this a Block Join, which clashes with the relational-database
terminology we have just used above.
\item Global-ordinals join -- it's an optimization of the base algorithm above,
but it has a limitation that both relations must be stored in a single index 
and the join keys must be stored in a single field\footnote{\url{https://www.elastic.co/docs/reference/elasticsearch/mapping-reference/parent-join\#_parent_join_restrictions}}.
This is not always possible in real-life scenarios. 
In this paper we'd rather keep entities in two separate flush-produced indices
 with externally provided $ID$, $FK$ arbitrary values.
\item The initial join in Apache Solr~\cite{lucidworks_solr_joins} implemented against
 inverted indices has some potential, but hardly able to compete in runtime with other ones.
\end{itemize}
There is also much research on distributed joins, which spread the
computation across a cluster of machines. We consider it out of scope, and
complementary rather than competing: a distributed algorithm still has to join
locally on each node, and the single-machine algorithm described in this paper
can serve as exactly that building block.

\subsection{The join-index lineage}
\label{sec:lineage}

Join indices were discussed in the relational-database field long ago:
\begin{itemize}
    \item Valduriez~\cite{valduriez1987} introduced the idea of a join index, which is a precomputed binary relation of surrogate pairs, with a read/write cost model showing when materialization beats query-time joining;
    \item O'Neil~\cite{oneil1997} introduced bitmap join indexes, which are industrial descendants of Valduriez's idea.
\end{itemize}
This paper adopts this idea for LSM-tree storages,
where search engines based on inverted index can be considered as an instance of it \S\ref{subsec:segments}.
The key challenge is to make join indices work in append-only flush-based storage, where the values ordinals
are segment scope and hard to correlate between segment and joining indices.

As the final note: we need to disambiguate \emph{join index} (materialized structure, this
paper) from \emph{index join} --- index nested-loops, using an index
during the join's inner loop \S\ref{sec:qtimejoin}.



\section{Problem Statement, Terminology, and Notation}
\label{sec:problem}


\subsection{Data Model}
\label{subsec:datamodel}

Let $\R$ be the \emph{parent} relation (the returned side, primary
key $\R.id$) and $\Srel$ the \emph{child} relation (the filtered
side, foreign key $\Srel.fk$), in a 1:M relationship. Let's call them just \emph{parent} and \emph{child}; and depict them as $\R$ and $\Srel$ in formal notation and pseudocode. (In Apache Solr terminology, child is the \texttt{from}
side and parent the \texttt{to} side.)

Each relation is stored as immutable, append-only segments:
$\R = R_1 \uplus \dots \uplus R_N$,
$\Srel = S_1 \uplus \dots \uplus S_M$; subscript $i$ always ranges
over parent segments, $j$ over child segments.
Inside a segment a
document is identified by its docid: $p \in [0, |R_i|)$,
$c \in [0, |S_j|)$.

\subsection{Filtered Semijoin}
\label{subsec:filteredsemijoin}

\begin{definition}[Filtered semijoin]
\label{def:filteredsemijoin}
Given arbitrary engine queries $q_R$ over $\R$ and $q_S$ over
$\Srel$, compute
\[
\mathit{Answer} \;=\; \sigma_{q_R}(\R) \ltimes \sigma_{q_S}(\Srel)
\;=\; \{\, r \in \R \mid r \models q_R \;\wedge\; \exists s \in
\Srel :\; s.fk = r.id \;\wedge\; s \models q_S \,\}.
\]
\end{definition}
Here $q_R$, $q_S$ denote arbitrary predicates over $\R$, $\Srel$
respectively; in SQL,

\texttt{SELECT * FROM R WHERE $q_R$ AND EXISTS (} \\
\texttt{\quad\quad\quad\quad SELECT 1 FROM S WHERE S.fk = R.id AND $q_S$)}.

\paragraph{Where $q_R$ is evaluated.}
The two predicates are not handled symmetrically, and it is worth saying so
before the algorithms, since $q_R$ appears in none of them. The algorithms of
\S\ref{sec:algorithm} compute only the $\ltimes$ operand --- the parents
having at least one child that matches $q_S$ --- and expose it as an ordinary
query over $\R$. The engine then evaluates
$\sigma_{q_R}(\R) \ltimes \sigma_{q_S}(\Srel)$ as the conjunction of $q_R$
with that query, exactly as it would intersect any two leaf queries; $q_R$ is
a sibling conjunct, never an input to the semijoin. Nothing is lost by this
split, and something is gained: because the semijoin is a conjunct rather
than a pipeline stage, $q_R$ can steer it --- restricting which parent
segments are visited at all (\S\ref{sec:segpruning}) and which parent docids
are ever offered for confirmation (\S\ref{subsec:lazytwophase}).

\subsection{Columns and Query Execution}
\label{subsec:columnsandquery}

A column (SortedDocValues field\footnote{\url{https://lucene.apache.org/core/10_1_0/core/org/apache/lucene/document/SortedDocValuesField.html}}) $f$ is a column of value ordinals,
positionally indexed by docid: $R_i.f[p]$ yields the value ordinal for
docid $p$, not the value itself. Since each field's dictionary is
private to its own segment, we annotate the value dereference with
segment (superscript) and field (subscript): the value is obtained
by a separate positional lookup on that ordinal,
$\texttt{values}^{R_i}_{f}[R_i.f[p]]$ (\S\ref{sec:valueordinals}). Square
brackets denote a positional column lookup.

Besides columns, a segment also exposes \emph{indexed fields}
(postings fields), searchable by value rather than by position. Given
a set of values $V$, a single \emph{bulk search} operation
\[
\mathit{search}(R_i.g, V) \;=\; \{\, p \in [0, |R_i|) \mid R_i.g[p]
\in V \,\}
\]
returns all matching docids at once, rather than one search per
value. For instance, searching a collected FK set against the
primary key field yields $\mathit{search}(R_i.id, \{FK\})$, the set
of parent docids $p$ whose $\mathit{id}$ matches some collected key
--- this is stage 2 of the classic query-time join
(\S\ref{sec:qtimejoin}).

Executing a query against a segment
yields a match iterator $\M(q, S_j) \subseteq [0, |S_j|)$
supporting $\mathit{next}()$, $\mathit{advance}(t)$, and the
two-phase pair $\mathit{approximation}()$/$\mathit{matches}(d)$.
Both $\mathit{next}()$ and $\mathit{advance}(t)$ return docids in ascending
order, and report exhaustion by returning the sentinel $\mathrm{EXHAUSTED}$.
We take $\mathrm{EXHAUSTED}$ to be greater than every docid, so that a bound
test such as $c \leq c^{ij}_{\max}$ fails of its own accord once the iterator
is spent and needs no separate exhaustion check; Lucene realizes the sentinel
as $\texttt{NO\_MORE\_DOCS} = \texttt{Integer.MAX\_VALUE}$, which has exactly
this property.

\subsection{Requirements and Result Sets}
\label{subsec:requirementsandresults}

Requirements, as set by
Sections~\ref{sec:intro}--\ref{sec:related}:
\begin{itemize}
    \item[R1.] performance: no query-time translation of external
               variable-length keys;
    \item[R2.] composability with arbitrary engine queries and
               their computation-skipping optimizations;
    \item[R3.] parallelism via the existing segment mechanism;
\end{itemize}

Result sets of the algorithm (all per parent segment): the exact
semijoin set $P_i$, which the confirmation phase computes in full; the
false-positive approximation $\Ahat{i} \supseteq P_i$; and the
\emph{half-read union} $H_i \subseteq P_i$, a false-negative set:
$p \in H_i \Rightarrow p \in P_i$, while
$p \notin H_i$ decides nothing until $H_i$ has converged to $P_i$.

\section{The Join Index}
\label{sec:joinindex}

\subsection{Formal Definition}
\label{sec:joinindex-formal}

The classic query-time join (\S\ref{sec:qtimejoin}) can be depicted
formally as a two-stage function from a set of child docids to, for
each parent segment, a set of matching parent docids. Given a set of
child docids $C \subseteq [0, |S_j|)$,
\begin{align*}
\mathit{QTJ}(C) &\;=\; (P_1, \dots, P_N), \\
V(C) &\;=\;
\underbrace{\bigl\{\, \texttt{values}^{S_j}_{fk}[S_j.fk[c]] \;\bigm|\; c \in C \,\bigr\}}_{\text{stage 1: collect FK values}}, \\
P_i &\;=\;
\underbrace{\mathit{search}\bigl(R_i.id,\, V(C)\bigr)}_{\text{stage 2: bulk search}}.
\end{align*}

We instead introduce, for every segment pair $(R_i, S_j)$, a
join-index column $\Ji$ mapping child docid directly to parent
docid, $\Ji[c] = p$:

\begin{definition}[Join-index column]
\label{def:joinindexcolumn}
For every pair $(R_i, S_j)$ we materialize an integer array $\Ji$ of
length $|S_j|$, indexed by child docid (on when it is written, see
\S\ref{sec:joinindex-impl}):
\[
\Ji[c] =
\begin{cases}
p, & \text{if the parent of } c \text{ lives in } R_i, \text{ i.e.\ } p
     \text{ is the unique docid of } [0, |R_i|) \text{ with}\\[2pt]
   & \quad \texttt{values}^{R_i}_{id}[R_i.id[p]] =
     \texttt{values}^{S_j}_{fk}[S_j.fk[c]],\\[4pt]
\bot, & \text{otherwise: no such } p \text{ --- the parent lives in
     another segment, or is absent.}
\end{cases}
\]
\end{definition}

The index to join $\Srel.fk = \R.id$ is thus a column of parent
docid indexed by child docid: $\Ji[c] = p$.

\paragraph{Well-definedness.}
Writing $\Ji[c] = p$ presupposes that $p$ is unique: a given foreign-key
value must select at most one parent docid, and in at most one parent
segment. That is stronger than $\R.id$ being a primary key of $R_i$ alone
--- it has to be a primary key of $\R$ as a whole. Search engines do not
enforce it the way a relational system does: in Apache Lucene an update is
a deletion followed by an append, so the same $id$ value is routinely
present in two segments at once, and duplicate $id$s are not rejected at
all. Two properties rescue the definition:
\begin{itemize}
    \item \emph{immutability}: a segment never changes once written, so
          $\Ji$ is a pure function of the pair $(R_i, S_j)$ --- once built
          it can never go stale, and the fresh copy of an updated parent
          simply lands in a new segment $R_k$ with its own column $J_{kj}$;
    \item \emph{deletion marks}: the obsolete copy is tombstoned, and is
          therefore excluded either when the column is built
          (\S\ref{sec:joinindex-impl}) or by the engine's live-docs filter
          when $P_i$ is scored.
\end{itemize}
So uniqueness is only required among \emph{live} documents, which is the
same condition the classic query-time join of \S\ref{sec:qtimejoin}
already relies on. Where even that fails --- genuinely duplicated live
$id$s --- $\Ji$ degenerates from a function into a relation, and the
algorithms of \S\ref{sec:algorithm} would report only one of the
duplicated parents per child, i.e.\ produce false negatives. We assume
unique live $id$s throughout.

\subsection{Implementation Notes}
\label{sec:joinindex-impl}

Here, \emph{indexed by child docid} means array access. In the Apache Lucene implementation, we
realize this as an \emph{auxiliary index}: a set of segments holding
\emph{join-index} columns  as NumericDocValues\footnote{\url{https://lucene.apache.org/core/10_1_0/core/org/apache/lucene/document/NumericDocValuesField.html}}, one column per segment
pair $(R_i, S_j)$, named accordingly. The docids of this auxiliary
index coincide with the docids of the child segment: writing
$U_j = [0, |S_j|)$ for that docid space, an auxiliary segment holding a
column for $S_j$ is addressed over $U_j$. ($U_j$ is the full address space
of $S_j$, not to be confused with the child docid \emph{set} $C$ of
\S\ref{sec:joinindex-formal}, which ranges over it.)

Join-index columns may be stored one per auxiliary index segment, or
several to a segment, since their names disambiguate them. Packing
multiple columns per segment reduces number of files open, but
complicates sweeping columns whose parent or child segment has since
been deleted.

This use of Lucene is unconventional: as a rule in the single segment 
same docid across different fields refers to the same document. 
But if we put several columns of different sizes into the single 
auxiliary index segment the columns of the different $j$ (child segments) are
projected onto one docid space although their $U_j$ differ,
 so a docid from one join-index column means nothing in another column.

We accept
this because we are piggybacking on existing Lucene machinery rather
than building bespoke storage.

\paragraph{When columns are materialized.}
Join-index columns are not written when the parent and child indices are
built. They are materialized on demand, by the first query that needs
them, and persisted thereafter: a segment pair is immutable
(\S\ref{sec:joinindex-formal}), so a column, once written, is valid for as
long as both its segments live, and every later query reads it rather than
rebuilding it.

Within a query, the work happens in bulk and up front rather than segment by
segment. When the join query builds its weight, it determines the full set of
pairs $(R_i, S_j)$ it will need, and writes every one of them that is not yet
present in a single batch, before scoring begins. Only pairs that this pass
misses are built lazily, per parent segment, inside the algorithm of
\S\ref{sec:algorithm}. We describe both because the fallback is what makes the
algorithm total --- it never has to fail on a missing column --- but it is the
bulk pass that does the work in a steady system, and the cost reported in
\S\ref{sec:experiments} is that pass, paid once, on the query path.

Building up front rather than on first access is a deliberate trade. It
forfeits the chance to skip columns that pruning would have made unnecessary
--- a pair whose column is never read still gets written --- in exchange for
one batched write per query instead of a write per parent segment, and for a
complete view of the join index at scoring time. Deferring the write to the
point of use, so that segment-level and a-priori pruning can suppress it
entirely, is left for future work (\S\ref{sec:future}).

We omit the algorithm for creating join-index columns: it is straightforward, and
the details may vary. The one we use can be seen in the actual implementation
(\S\ref{sec:implementation}).

It is worth noting that building a join-index column honours the deletion marks on
both sides, which makes the column smaller, since fewer values enter it.

Each join index column additionally receives approximation ranges: the min/max child docid
it is defined at, and the min/max parent docid it maps onto,
\[
c^{ij}_{\min} = \min \{\, c \mid \Ji[c] \neq \bot \,\}, \qquad
c^{ij}_{\max} = \max \{\, c \mid \Ji[c] \neq \bot \,\},
\]
\[
p^{ij}_{\min} = \min \{\, p \mid \exists\, c : \Ji[c] = p \,\}, \qquad
p^{ij}_{\max} = \max \{\, p \mid \exists\, c : \Ji[c] = p \,\}.
\]

\paragraph{Empty columns.}
A segment pair may have no join at all: if no child of $S_j$ has its parent in
$R_i$, then $\Ji$ is $\bot$ everywhere, the four sets above are empty, and the
bounds are undefined as written. This is not a corner case to be waved away. It
is the \emph{common} case exactly when the two sides' insertion orders are
correlated --- a bulk load ordered so that a parent lands near its children
leaves most of the $N \times M$ pairs disjoint --- which is the regime in which
the ranges are worth anything at all
(\S\ref{sec:pruninginstrumentation}). We therefore take the conventional
extrema over the empty set,
\[
c^{ij}_{\min} = p^{ij}_{\min} = +\infty, \qquad
c^{ij}_{\max} = p^{ij}_{\max} = -\infty,
\]
realized as $\mathrm{EXHAUSTED}$ and $-1$ respectively. Every guard in
\S\ref{sec:algorithm} then bypasses such a column of its own accord, in the same
way and for the same reason that the sentinel handles a spent iterator
(\S\ref{subsec:columnsandquery}): $\mathit{advance}(+\infty)$ returns
$\mathrm{EXHAUSTED}$, the bound test $c \leq c^{ij}_{\max}$ fails for every $c$,
and $[p^{ij}_{\min}, p^{ij}_{\max}]$ is the empty interval, contributing nothing
to $\Ahat{i}$. No algorithm needs a case for it.

An empty column consequently need not be materialized: its metadata alone
answers every query that would have read it. The prototype detects the case
before building, by testing whether the $\R.id$ and $\Srel.fk$ term ranges of
$(R_i, S_j)$ are disjoint --- a sufficient, not necessary, condition --- and in
that case writes the metadata only. What this does \emph{not} do is reduce the
$N \times M$ column count to address and sweep
(\S\ref{sec:materializationcost}); it removes the bytes, not the bookkeeping.

Note: below we approximate the set of docids by a range with false-positive entries.
 In future work we may extend this to $n$-range approximation.

There is a merge policy for the auxiliary join index which marks a whole segment
(all its columns) for deletion once every segment pair $(R_i, S_j)$ it holds a
column for has been deleted from the parent index $\R$ and the child index
$\Srel$.


\section{The Semijoin Algorithm}
\label{sec:algorithm}

\subsection{Base algorithm: eager per-segment semijoin}
\label{sec:base}

We reuse Lucene's built-in parallel segment search, giving
parent-segment concurrency for free. For every parent segment we
read the join-index columns of all child segments:
\[
P_i \;=\; \bigcup_{j=1}^{M} \bigl\{\, \Ji[c] \;\bigm|\;
c \in \M(q_S, S_j),\; \Ji[c] \neq \bot \,\bigr\}.
\]

Thus we calculate the semijoin of a parent segment from child
predicate matches, in parallel across parent segments, with no barrier between
stages at which all parent tasks must have arrived.

\begin{algorithm}[H]
\caption{Eager per-segment semijoin (base algorithm)}
\label{alg:base}
\begin{algorithmic}[1]
\REQUIRE parent segments $R_1..R_N$, child segments $S_1..S_M$,
         join-index columns $\{\Ji\}$, child query $q_S$, $c^{ij}_{\min}$, $c^{ij}_{\max}$ - ranges of child docids for each join-index column
\REQUIRE $\mathit{cursor}(j)$: a fresh cursor over $B_j$, the shared cached
         bitset of $\M(q_S, S_j)$. The first caller for a given $j$ ---
         whichever parent task that is --- evaluates $q_S$ over $S_j$ to
         exhaustion and publishes $B_j$ under a per-$j$ latch; every later
         caller receives a cursor over it. Shared across all parent tasks;
         see the cost discussion below
\ENSURE per-segment semijoin sets $P_i$, emitted as sorted docids
        iterators over $R_i$
\FOR{\textbf{each} parent segment $R_i$ \textbf{in parallel}}
    \STATE $P_i \leftarrow \emptyset$ \COMMENT{e.g.\ a bitset of $|R_i|$ bits}
    \FOR{$j \leftarrow 1$ \TO $M$ \COMMENT{parallelizable too, but kept sequential; see below}}
        \STATE $it \leftarrow \mathit{cursor}(j)$ \COMMENT{sorted match cursor; $q_S$ is retrieved over $S_j$ at most once overall}
        \STATE $c \leftarrow it.\mathit{advance}(c^{ij}_{\min})$ \COMMENT{skip below child range}
        \IF{$c > c^{ij}_{\max}$}
            \STATE \textbf{continue} \COMMENT{a-priori column bypass, \S\ref{sec:apriori}}
        \ENDIF
        \WHILE{$c \neq \mathrm{EXHAUSTED}$ \AND $c \leq c^{ij}_{\max}$}
            \STATE $p \leftarrow \Ji[c]$ \COMMENT{single array lookup}
            \IF{$p \neq \bot$}
                \STATE $P_i \leftarrow P_i \cup \{p\}$
            \ENDIF
            \STATE $c \leftarrow it.\mathit{next}()$
        \ENDWHILE
    \ENDFOR
\ENDFOR
\end{algorithmic}
\end{algorithm}

Note that the inner loop over child segments is parallelizable as well: every
iteration reads its own join-index column $\Ji$ and its own child segment, so
the iterations are independent except for the accumulator $P_i$ they share. We
deliberately do not pursue this opportunity. The outer loop already yields $N$
independent tasks --- for free, via Lucene's parallel segment search --- which
is normally enough to saturate the available cores, whereas a second, nested
level of concurrency would have several threads updating a single $P_i$. That
would demand either synchronized updates or per-thread bitsets merged
afterwards, reintroducing exactly the sequential merge step whose cost we
criticize in the classic query-time join
(\S\ref{sec:qtimejoin}). We therefore keep child segments sequential
within each parent segment.

\paragraph{Cost: the child query is evaluated $N \times M$ times.}
The match iterator $\M(q_S, S_j)$ is constructed inside the parent-segment
loop, so the child query is evaluated once per segment pair: $N \times M$
evaluations in total, or $N$ times over each child segment $S_j$. The
classic query-time join (\S\ref{sec:qtimejoin}) evaluates it $M$ times --- once
per child segment --- and this redundancy is precisely the price we pay for
dropping its synchronization point. We trade one sequential merge of $\{FK\}$
for $N$ independent re-enumerations that need not coordinate at all. Two
things make that trade favourable.

First, the child query is cheap relative to what the classic algorithm does
with its matches. There, every child match yields an external, variable-length
FK \emph{value}, all such values are deduplicated into one global set across
all child segments --- the sequential step --- and that set is then bulk-searched
against $R_i.id$. Here, a child match costs a single array lookup
$\Ji[c]$ (\S\ref{sec:joinindex-formal}), with no key materialization and no
dictionary lookups on either side; and the pruning hierarchy of
\S\ref{sec:pruning} removes much of even that, since a column whose child
range does not intersect the matches is bypassed without being loaded at all,
while the lazy variant of \S\ref{subsec:lazytwophase} stops as soon as the
parent docid under confirmation is found.

Second, the repetition is enumeration, not retrieval. The child matches of a
segment are materialized once, globally, and then shared --- this is the
$\mathit{cursor}(j)$ of Algorithm~\ref{alg:base}, and it is load-bearing enough
for the cost argument that we make it an explicit input rather than leaving it
to the implementation. The first parent task
to touch $S_j$ --- whichever one that happens to be --- evaluates
$\M(q_S, S_j)$ to exhaustion and publishes the resulting bitset
$B_j = \{\, c \in \M(q_S, S_j) \,\}$; every later toucher receives
a cursor over $B_j$. The cache is thus triggered lazily but filled
eagerly, and the persisted $it_j$ of Algorithm~\ref{alg:lazyhalfread} are $N$
independent cursors over one shared bitset rather than $N$ independent
executions of $q_S$. Retrieval --- postings decoding, skipping, and the
evaluation of $q_S$'s own subqueries --- therefore happens at most $M$ times in
total, once per child segment, and it is only the enumeration of an
already-materialized bitset that is repeated up to $N$ times per segment.
Note that this cache does not reinstate the synchronization point of the
classic query-time join (\S\ref{sec:qtimejoin}). Two tasks that reach an
uncached $S_j$ at the same moment resolve it as any concurrent memoization
does: one fills, the other waits on that single fill. The parent tasks never
wait on each other's \emph{results} --- there is nothing to merge, and no point
at which all $N$ of them must have arrived --- only, at worst, on the first
materialization of a scan they both need anyway.

\paragraph{What this leaves for the pruning hierarchy to save.}
It follows that the levels of \S\ref{sec:pruning} save on the $N \times M$ side
of the computation rather than on the $M$ side, and it is worth being exact
about which reads they eliminate. An early exit in
Algorithm~\ref{alg:lazyhalfread}, and an a-priori bypass in any of the three variants,
avoid reading the join-index column $\Ji$ --- one of $N \times M$ objects, each
of which has to be located, loaded and decoded --- but they do not avoid the
child scan underneath it, which a sibling parent task may have paid for
already, or will pay for later. When the $R_1$ trace of
\S\ref{app:example-lazy} reports that $J_{13}$ is never read, the claim is
about that column and that parent segment: $S_3$'s matches are materialized as
soon as any parent task reaches $S_3$. This is the side worth saving on,
because it is the side that grows with $N$.

\subsection{Two-phase semijoin: approximate, then confirm}
\label{sec:twophase}

Here \emph{two-phase} carries its Lucene meaning throughout: the
approximation/confirmation iterator contract of \S\ref{sec:dynamicpruning},
exposed as the $\mathit{approximation}()$/$\mathit{matches}(d)$ pair of
\S\ref{subsec:columnsandquery}. It is unrelated to the two \emph{stages} of
the classic query-time join (\S\ref{sec:qtimejoin}); what follows is our
algorithm implementing that contract, not a variant of that one.

Every join-index column additionally has parent docids range:
$p^{ij}_{\min}, p^{ij}_{\max}$; leaping over all child
segments (those that passed segment-level pruning,
\S\ref{sec:segpruning}) and unioning all parent docid ranges,
we get a false-positive match approximation:
\[
\Ahat{i} \;=\; \bigcup_{j=1}^{M}
\bigl[\, p^{ij}_{\min},\; p^{ij}_{\max} \,\bigr]
\;\supseteq\; P_i .
\]
The advantage is that it yields parents approximation (with false positives) even without reading join-index column, thus such candidates might be withdrawn by \S\ref{sec:dynamicpruning} or, for example, intersecting with highly restrictive parents filter.

This idea can be extended to $n$ range edges, approximating
$\{p\}$ as a union of ranges, although it has not been evaluated.

Then, to confirm a potential parent docid match, we need to union
all confirmations $\{p\}$ per child segment: for each child
segment, the corresponding index column is read at the child
matches, collecting $\{p\}$, and then unioned per child segment;
thus we get the true-positive parent matches for the current
parent segment --- which is the exact semijoin set $P_i$ of
\S\ref{sec:base} again, now computed on demand rather than eagerly.

\begin{algorithm}[H]
\caption{Two-phase semijoin (approximate, then confirm)}
\label{alg:twophase_alg}
\begin{algorithmic}[1]
\STATE $\mathit{approximation}(R_i)$: \textbf{return}
       $\Ahat{i} = \bigcup_j [p^{ij}_{\min}, p^{ij}_{\max}]$
       \COMMENT{metadata only; no column reads}
\STATE $\mathit{matches}(R_i)$:
       $P_i \leftarrow \bigcup_j \{ \Ji[c] : c \in \M(q_S, S_j) \cap
       [c^{ij}_{\min}, c^{ij}_{\max}] \}
       \setminus \{\bot\}$ \COMMENT{full read, as in
       Algorithm~\ref{alg:base}}
\STATE answer membership of $p \in \Ahat{i}$ by $p \in P_i$
\end{algorithmic}
\end{algorithm}

Implementation detail: this refined parent doc set $P_i$
may replace the approximation docset for following processing,
although it might not be fully compatible with possible Lucene
optimizations.

\subsection{Lazy confirmation: the half-read union}
\label{subsec:lazytwophase}

Then, we can make it even lazier. 

Let us run only a few per-child-segment steps of
Algorithm~\ref{alg:twophase_alg}, until we confirm
the single given parent docid --- and break. Such a half-read union is called a
\emph{false-negative doc set} $H_i$: a one in it is a match for sure,
while a zero is uncertain, since it is a half-read union.

For the next parent docid arriving for confirmation, we can check
the false-negative set first. Note that only false-positive
documents arrive for confirmation; and only if the false-negative
check does not confirm the match do we proceed with the following
step of the lazy variant, dumping the next $\{p\}$ into
the false-negative set, until it processes all child segments and
converges to the two-phase algorithm above, updating the underlying
match approximations with only true matches.

\begin{algorithm}[H]
\caption{Lazy confirmation via the half-read union}
\label{alg:lazyhalfread}
\begin{algorithmic}[1]
\REQUIRE parent segment $R_i$; child query $q_S$; join-index columns
         $\Ji$ with metadata $c^{ij}_{\min}, c^{ij}_{\max}$, $j = 1..M$;
         $\mathit{cursor}(j)$ as in Algorithm~\ref{alg:base}
\REQUIRE $\mathit{step}(j) \equiv$ on the first use of $S_j$:
         $it_j \leftarrow \mathit{cursor}(j)$, then
         $it_j.\mathit{advance}(c^{ij}_{\min})$; $it_j.\mathit{next}()$ thereafter
         \COMMENT{a-priori lower bound, \S\ref{sec:apriori}}
\ENSURE  $\mathit{matches}(p)$ decides $p \in P_i$ for each candidate $p$
         offered by $\Ahat{i}$
\STATE \textbf{initialize} once per parent segment $R_i$, then persist
       across calls:
\STATE \quad $H_i \leftarrow \emptyset$
\STATE \quad $\mathit{unread} \leftarrow$ queue of child segments $1..M$
\STATE \quad $it_j \leftarrow$ \textbf{undefined} for each $j$
       \COMMENT{cursors are created by $\mathit{step}$, on first touch}
\STATE \textbf{function} $\mathit{matches}(p)$:
\IF{$p \in H_i$}
    \RETURN \TRUE \COMMENT{false-negative check first --- free}
\ENDIF
\WHILE{$\mathit{unread} \neq \emptyset$}
    \STATE $j \leftarrow \mathit{unread}.\mathit{head}$
    \WHILE{$(c \leftarrow \mathit{step}(j)) \neq \mathrm{EXHAUSTED}$
           \AND $c \leq c^{ij}_{\max}$}
        \STATE $p' \leftarrow \Ji[c]$ \COMMENT{first read of $\Ji$ loads the column}
        \IF{$p' \neq \bot$}
            \STATE $H_i \leftarrow H_i \cup \{p'\}$
            \IF{$p' = p$}
                \RETURN \TRUE \COMMENT{confirmed --- break early}
            \ENDIF
        \ENDIF
    \ENDWHILE
    \STATE $\mathit{unread}.\mathit{pop}()$
    \COMMENT{$S_j$ drained, or past $c^{ij}_{\max}$}
\ENDWHILE
\RETURN \FALSE \COMMENT{$H_i$ has converged: $H_i = P_i$}
\end{algorithmic}
\end{algorithm}

The two guards on the inner loop are the a-priori level of
\S\ref{sec:apriori} composed into the document level, not an alternative to
it: $\mathit{step}$ skips the child matches below $c^{ij}_{\min}$ on the first
touch of $S_j$, and the $c \leq c^{ij}_{\max}$ bound retires the segment as
soon as its matches run past the column's range. They subsume the column
bypass as a degenerate case --- if the first stepped match already exceeds
$c^{ij}_{\max}$, the loop body never executes, $S_j$ is popped, and $\Ji$ is
never loaded --- so a column outside the child query's range costs one
$\mathit{advance}$ and no column read at all. Algorithm~\ref{alg:base} and the
$\mathit{matches}$ line of Algorithm~\ref{alg:twophase_alg} apply the same
restriction, the latter by intersecting the child matches with
$[c^{ij}_{\min}, c^{ij}_{\max}]$; the three levels of \S\ref{sec:pruning} are
therefore stacked in every variant, not distributed among them. An all-$\bot$
column needs no case of its own here either: under the conventions of
\S\ref{sec:joinindex-impl} its $c^{ij}_{\min}$ is $+\infty$, so the first
$\mathit{step}$ returns $\mathrm{EXHAUSTED}$, $S_j$ is popped unread, and the
column --- which was never materialized --- is never asked for.

Note that $\mathit{step}$ is also what \emph{creates} $it_j$, on first touch,
rather than the initialization block doing so for all $j$ at once. The
distinction is not cosmetic: constructing a cursor is what triggers the
materialization of $B_j$ (\S\ref{sec:base}), so creating all $M$ of them up
front would make every parent segment pay for every child segment's scan before
its first $\mathit{matches}$ call, and would leave the early exits of this
algorithm saving column reads only. Deferring creation to first touch is what
lets a parent segment that converges early --- the $R_1$ trace of
\S\ref{app:example-lazy} --- avoid $S_3$ entirely when no sibling parent task
has reached it.

\paragraph{When laziness pays, and when it does not.}
The asymmetry of $H_i$ has a sharp consequence that we state plainly, since it
bounds what document-level pruning can deliver. A confirmation is cheap: the
loop stops at the first $c$ with $\Ji[c] = p$. A \emph{refutation} is not:
$H_i$ is a false-negative set, so $p \notin H_i$ decides nothing, and
Algorithm~\ref{alg:lazyhalfread} may only answer \textsc{false} once $\mathit{unread}$
is empty --- that is, once $H_i$ has converged to $P_i$. The first parent
docid that is not a match therefore pays the full cost of the eager algorithm,
and the lazy variant degenerates to Algorithm~\ref{alg:base} at that moment.
Everything after it is free, as the $R_2$ trace of \S\ref{app:example-lazy}
shows, but nothing before it was saved.

Whether that moment arrives early depends on how tight $\Ahat{i}$ is, and with
a single range per column it is typically not tight at all. Any join-index
column whose parents are spread over $R_i$ has $p^{ij}_{\min}$ near $0$ and
$p^{ij}_{\max}$ near $|R_i|-1$, so the union in \S\ref{sec:twophase} covers
almost the whole segment --- as it does for both segments of the worked
example (\S\ref{app:example-approx}). Every non-matching parent docid that the
sibling conjunction admits is then a false positive, and the first of them
triggers convergence. Document-level pruning consequently pays off in the
regime where the parent-side filter is restrictive and the candidates
surviving it are mostly true matches: there, confirmations dominate,
refutations are rare, and whole child segments are never touched. It pays off
least when candidates are drawn from a broad $\Ahat{i}$, which is precisely
when the approximation was uninformative to begin with.

This is the strongest argument for the $n$-range extension sketched in
\S\ref{sec:twophase} and \S\ref{sec:apriori}: approximating $\{p\}$ by a union
of disjoint ranges rather than a single interval shrinks $\Ahat{i}$, which
suppresses exactly the false positives that force early convergence. We
therefore regard it not as an optional refinement but as what makes the
document level robust across data distributions; it remains unevaluated.

\paragraph{Ordering the child segments.}
The order in which $\mathit{unread}$ is consumed matters for the same reason.
Ordering child segments by descending cardinality of their join-index column,
$\bigl|\{\, c \mid \Ji[c] \neq \bot \,\}\bigr|$, visits first the columns that
cover the most parents and are hence the likeliest to contain the docid being
confirmed, so a confirmation is expected to be reached after fewer reads. The
qualification is the one just made: this is a heuristic for the confirmation
path alone. A refutation drains every column whatever the order, and costs the
same under every permutation --- reordering redistributes work inside the
early-exit case and leaves the worst case exactly where it was.

Two practical notes. The cardinality has to be recorded per column, alongside
the bounding ranges of \S\ref{sec:joinindex-impl}; the prototype stores the
ranges but not yet the cardinality, so this ordering is left for future work
and has not been evaluated. And because the order is a per-task heuristic,
parent tasks running concurrently may reach the child segments in different
orders, which leaves it undetermined in advance which task pays a given child
segment's first-touch materialization (\S\ref{sec:base}). That cost is paid
once either way, so the indeterminacy is one of accounting rather than of
total work.

\section{Pruning Hierarchy}
\label{sec:pruning}

We now summarize the pruning hierarchy of the semijoin algorithm described above.

\subsection{Segment-level pruning}
\label{sec:segpruning}

If the
semijoin query is intersected (AND) with  another query which does
not have matches in a certain parent segment, that segment will be
completely skipped from processing --- which is more efficient than
the classic query-time join (\S\ref{sec:qtimejoin}), which calculates FKs that can be dropped
by the parent-level filter.

This level is worth stating precisely, because we do not implement it: it
falls out of the algorithm's shape. Because the semijoin is an ordinary
per-segment query rather than a global two-stage pipeline, segment-level
skipping comes for free --- the engine simply never asks $R_i$ for a scorer
when a sibling conjunct has no matches there, exactly as it would for any
other leaf query. Algorithm~\ref{alg:base} has no line corresponding to it.
The contrast with the classic query-time join (\S\ref{sec:qtimejoin}) is
therefore structural rather than a matter of a cleverer test: its stage~1
collects FKs from \emph{all} child segments before any parent-side filter is
consulted, so work spent on children whose parents a sibling filter will
discard has already been spent by the time the filter applies. Being
per-segment is what makes the saving available; the two levels below are what
the algorithm does with it.

\subsection{A-priori (min/max) pruning}
\label{sec:apriori}

Another optimization is a-priori pruning: in addition to the index
column, we also write a kind of bounding box over join-index column $p^{ij}_{\min}, p^{ij}_{\max}$ and
$c^{ij}_{\min}, c^{ij}_{\max}$.

The child docid range $c^{ij}_{\min}, c^{ij}_{\max}$ lets us:
\begin{itemize}
    \item skip the first child matches with $c < c^{ij}_{\min}$, then
    \item prune at the first child filter match: if $c(\text{first match}) > c^{ij}_{\max}$, loading this column can be bypassed completely. Then, 
    \item when we loop through child matches, it can exit earlier when $c > c^{ij}_{\max}$ 
\end{itemize}

The parent docid ranges $p^{ij}_{\min}, p^{ij}_{\max}$  lets us build an approximation of the parent matches without reading the index column at all, which is used in the two-phase algorithm \S\ref{sec:twophase}.

These ideas might be extended over $n$ even edge values
approximating the set of docids by disjoint ranges (interval sketches) and allowing leapfrogging
of child matches when calculating the semijoin, and narrow down parents approximation --- although the
efficiency of this optimization depends on the actual data and has
not been evaluated. 

\subsection{Document-level pruning}
\label{sec:docpruning}

Document-level pruning is the lazy two-phase machinery of
Section~\ref{subsec:lazytwophase}: it defers loading index columns until it's inevitable, and even more, it can confirm the given parent docid by reading only part of the join-index column. That returns control over the iteration to potentially faster sibling queries and advances overall search faster. See
 \S\ref{sec:dynamicpruning}.
Unlike the two levels above, this one is not unconditional: it saves work on
\emph{confirmations} only, and the first candidate it cannot confirm costs a
full read of every remaining column. Its benefit is therefore
distribution-dependent, and is largest when the parent-side filter is
restrictive enough that most surviving candidates are true matches --- see the
discussion at the end of \S\ref{subsec:lazytwophase}.


\section{Implementation Notes}
\label{sec:implementation}

Thus we get an algorithm joining two flush-based tables with
external string variable-length keys. It does it concurrently: it
employs the built-in Lucene mechanism. It implements the a-priori and
document-level pruning of \S\ref{sec:apriori} and \S\ref{sec:docpruning}, and
inherits segment-level skipping from being an ordinary per-segment query
(\S\ref{sec:segpruning}), which together allow combining it with all Lucene
queries efficiently, without
wasting computation on matches which would later be thrown away by
another filter.

It uses Lucene data structures, which are compact, efficient, and
flexible: for example, we can trade off memory instead of disk
writes by switching to a memory buffer directory implementation.

The code is published as early draft at \url{https://github.com/apache/solr/compare/main...mkhludnev:solr:aijoin-qparser} under ASLv2. It requires no changes to Lucene itself, being client code that merely uses the library. Any authors of custom search servers may adopt this code by copying it.  

It also provides an Apache Solr query parser integration, which makes it readily adoptable by Solr users. For the codename of this query parser we use \texttt{aijoin} (``AI'' stands for ``auxiliary index'').

We defer Elasticsearch and OpenSearch integrations since they don't favor cross-index joins; however, it should be easy to integrate into a single-index join flow.

\section{Experimental Evaluation}
\label{sec:experiments}
\subsection{Existing Benchmark}

The benchmark is published at \url{https://github.com/mkhludnev/aijoin-benchmark}.
Sample benchmark results are reproduced in Table~\ref{tab:benchmark-results}.

The benchmark runs against bare Solr setup extended with the query parser \S\ref{sec:implementation}. 
It runs on cloud VM with 4 vCPUs, 8G RAM, 2G heap, and SSD storage.
Also, segment-level parallelism was set to 4 as described in \texttt{README.md}.

The indexer creates segmented index of 1M parent documents (products) and 10M
child documents (skus) in a single shard (horizontal partition in Solr terms).
The flush produced $N = 6$ parent segments and $M = 9$ child segments, hence
$N \times M = 54$ join-index columns; segment sizes are uneven, as
flush-produced runs are, the largest child segment holding 3{,}909{,}575 docs.
Every figure below scales with these two counts, so we state them once here.

The searcher then sends semijoin queries joining skus to products, with random
filters applied on both sides.

Three properties of the harness make the two parsers comparable, and all three
are deliberate. First, each arm runs a \emph{fixed number} of queries --- 500
--- rather than for a fixed duration, and the query list is generated up front
from a fixed seed, so query $k$ is the same query in every arm and at every
concurrency. Second, each arm runs at \emph{fixed concurrency} rather than at a
fixed request rate: a rate above what the server can absorb makes the queue
grow without bound, so the run would report queueing rather than joining, and
whether a given rate is sustainable differs between the two parsers by roughly
the factor we are trying to measure. A closed loop cannot diverge; we report
the rate each arm \emph{achieved} as a result rather than imposing one. Third,
we sweep the concurrency, because a single operating point cannot distinguish a
cheaper query from one that scales better.

\begin{table}[H]
\centering
\begin{tabular}{@{}lrrr@{}}
\toprule
\texttt{QTime}, ms & \texttt{\{!join score=none\}} & \texttt{\{!aijoin\}} & speedup \\
(concurrency 1)    & (baseline) & (join index) & \\
\midrule
min                         & 1273     & 187      & 6.8$\times$   \\
median                      & 1953     & 345      & 5.7$\times$   \\
p90                         & 2585     & 492      & 5.3$\times$   \\
p95                         & 2640     & 496      & 5.3$\times$   \\
p99                         & 2716     & 502      & 5.4$\times$   \\
mean                        & 1934.6   & 359.8    & 5.4$\times$   \\
\midrule
max                         & 3006     & 12188$^\dagger$ &          \\
\bottomrule
\end{tabular}
\caption{Per-query latency at concurrency one, 500 queries per arm, same query
set in both. $^\dagger$the join-index arm ran against a cold auxiliary index:
its slowest query is query~1 of 500, which paid the whole materialization of
\S\ref{sec:materializationcost}; the next slowest is 571\,ms, and excluding it
the mean is 336.1\,ms.}
\label{tab:benchmark-results}
\end{table}

\begin{table}[H]
\centering
\begin{tabular}{@{}lrrrr@{}}
\toprule
Concurrency & 1 & 2 & 4 & 8 \\
\midrule
\multicolumn{5}{@{}l}{\texttt{\{!join score=none\}} (baseline)} \\
\quad mean \texttt{QTime}, ms   & 1934.6 & 1962.0 & 3249.2 & 7027.7 \\
\quad p99 \texttt{QTime}, ms    & 2716   & 2710   & 4456   & 9576   \\
\quad throughput, queries/s     & 0.48   & 0.95   & 1.18   & 1.11   \\
\quad \emph{gain over previous} & ---    & 1.97$\times$ & 1.24$\times$ & 0.94$\times$ \\
\midrule
\multicolumn{5}{@{}l}{\texttt{\{!aijoin\}} (join index)} \\
\quad mean \texttt{QTime}, ms   & 359.8$^\dagger$ & 325.7 & 438.6 & 845.9 \\
\quad p99 \texttt{QTime}, ms    & 502    & 523   & 711   & 1478  \\
\quad throughput, queries/s     & 2.03$^\dagger$  & 4.43  & 7.09  & 8.04 \\
\quad \emph{gain over previous} & ---    & 2.18$\times$ & 1.60$\times$ & 1.13$\times$ \\
\midrule
Speedup, mean \texttt{QTime}    & 5.4$\times$ & 6.0$\times$ & 7.4$\times$ & 8.3$\times$ \\
Throughput ratio                & 4.2$\times$ & 4.7$\times$ & 6.0$\times$ & 7.2$\times$ \\
\texttt{numFound} mismatches    & 0/500 & 0/500 & 0/500 & 0/500 \\
\bottomrule
\end{tabular}
\caption{Concurrency sweep on 4 vCPUs, 500 queries per cell.
$^\dagger$cold auxiliary index, see Table~\ref{tab:benchmark-results}.
The baseline peaks at concurrency 4 and \emph{loses} throughput at 8; the join
index is still gaining there, at 6.8$\times$ the baseline's best figure at any
concurrency (the 7.2$\times$ in the row above is the two arms compared at
concurrency 8).}
\label{tab:concurrency-sweep}
\end{table}

\paragraph{Result-set equivalence.}
Because both arms execute the same 500 queries, they can be compared query by
query rather than in aggregate. They agree on \texttt{numFound} for every one
of the 500 --- zero mismatches --- which is what establishes that the pruning
of \S\ref{sec:pruning} trades no correctness for speed. Matching aggregate
distributions would not have established it: two runs of different lengths
execute different query sets, so agreement between their summary statistics is
neither necessary nor sufficient.

\paragraph{The gain is distributional, not an artifact of the mean.}
At concurrency one the join index is faster at every quantile we measured, by
between $5.3\times$ and $6.8\times$ from the minimum through the 99th
percentile. Its tail is also tighter in absolute terms: the spread from median
to p99 is 157\,ms against the baseline's 763\,ms.

\paragraph{The advantage widens with load, and that is the more informative result.}
Table~\ref{tab:concurrency-sweep} sweeps the concurrency on the same 4 vCPUs.
The baseline converts the first doubling into throughput almost perfectly
($0.48 \to 0.95$ queries/s, $1.97\times$), then falters ($1.24\times$), then
\emph{regresses}: at concurrency 8 it serves 1.11 queries/s, fewer than the
1.18 it managed at 4, while its mean latency more than doubles to 7027.7\,ms
and its p99 reaches 9576\,ms. Its peak is at concurrency 4, and past that point
added load costs throughput rather than buying it.

The join index does not turn over anywhere in the range we measured. It gains
$1.60\times$ from 2 to 4 and is still gaining at 8, where it serves 8.04
queries/s --- $6.8\times$ the baseline's best figure at any concurrency, which is
the sterner comparison than the $7.2\times$ separating the two arms at
concurrency 8. Its returns are diminishing there
($1.13\times$ for a doubling), so its own ceiling is nearby; we simply did not
reach it.

Normalizing removes the objection that a cheaper query would reach a higher
ceiling anyway. Over the identical step from concurrency 2 to 4 the join index
converts $80\%$ of the added concurrency into throughput against the baseline's
$62\%$; over the step from 4 to 8 it converts $56\%$ while the baseline
converts none. The two are not separated by a constant factor: they scale
differently, and the measured speedup rises monotonically with load ---
$5.4\times$, $6.0\times$, $7.4\times$, $8.3\times$.

A saturated system should hold its throughput, not shed it, so the baseline's
$6\%$ decline from 4 to 8 is contention rather than mere saturation. We did not
instrument it, but the shape of the classic algorithm suggests where to look:
it materializes a global $\{FK\}$ set per query (\S\ref{sec:qtimejoin}), sized
by the number of matching child documents, so eight in-flight queries hold
eight such sets at once on a 2\,GB heap. The join-index algorithm allocates
per-segment bitsets instead, whose size depends on segment length rather than
on match count. Confirming that this is what collapses the baseline needs
allocation and GC instrumentation we have not added.

This is the first direct evidence for requirement R3
(\S\ref{subsec:requirementsandresults}) and for the Amdahl argument of
\S\ref{sec:qtimejoin}: an algorithm with a sequential merge between its two
stages should reach its ceiling sooner than one without a synchronization
point, and it does --- at half the concurrency and at $1/6$ the throughput. We
state the limit of that inference too. The classic join also costs more CPU per
query --- at saturation the throughput ratio is essentially the ratio of CPU
work per query --- and a CPU-heavier query saturates a fixed core count earlier
whether or not it contains a serial section. The measurement is consistent with
the sequential merge being the cause; it does not isolate it. Doing so would
need the $\{FK\}$ merge instrumented separately, which we have not done.

\paragraph{Throughput ratios are smaller than latency ratios, for a client-side reason.}
Each query carries a client and network cost of roughly 120\,ms (median client
latency exceeds median \texttt{QTime} by 116\,ms in the join-index arm and
123\,ms in the baseline). Being constant in absolute terms, it dilutes a ratio
far more when the server-side figure is 345\,ms than when it is 1953\,ms, which
is why the throughput ratios in Table~\ref{tab:concurrency-sweep} run below the
\texttt{QTime} speedups. The speedups are server-side results; what an
application sees depends on what else is in its request path.

\paragraph{The materialization cost is one query, and it is visible.}
The concurrency-one join-index run started from a cold auxiliary index, and the
whole build lands on query~1 of 500: a \texttt{QTime} of 12188\,ms, against
571\,ms for the next slowest. That figure agrees with the 11325\,ms measured
inside the builder (\S\ref{sec:materializationcost}), which confirms that
\texttt{QTime} accounts for the build rather than hiding it. Spread over 500
queries it raises the arm's mean by 23.7\,ms, from 336.1 to 359.8; every other
cell of Table~\ref{tab:concurrency-sweep} ran warm. Moving that one query's
cost into searcher warming is future work (\S\ref{sec:future}).

\subsection{The cost of materialization}
\label{sec:materializationcost}

A join index is a materialized structure, so the question Valduriez's cost
model asks of it (\S\ref{sec:lineage}) is when materializing beats joining at
query time. This benchmark answers it directly. Building the auxiliary index
from cold produced the single event in
Table~\ref{tab:build-cost}; both figures it needs --- what the structure costs
and what it saves per query --- are now measured on the same workload.

\begin{table}[H]
\centering
\begin{tabular}{@{}lr@{}}
\toprule
Parent segments $N$ / child segments $M$    & 6 / 9 \\
Join-index columns $\Ji$ built              & 54 \\
Largest child segment $|S_j|$               & 3{,}909{,}575 \\
Non-$\bot$ entries over all columns         & 10{,}000{,}000 \\
Auxiliary index on disk                     & $\approx$30\,MB \\
Build time (one batch, one thread)          & 11{,}325\,ms \\
\midrule
Latency saved per query (\S\ref{sec:experiments}) & 1{,}574.8\,ms \\
Queries to break even                       & 7.2 \\
\bottomrule
\end{tabular}
\caption{Materialization cost, from a cold auxiliary index, against the
per-query saving of Table~\ref{tab:benchmark-results}.}
\label{tab:build-cost}
\end{table}

\paragraph{Break-even is seven queries.}
The build costs 11.3\,s once; each query thereafter is 1{,}574.8\,ms cheaper
than the classic query-time join at concurrency one, and more than that under
load. The structure has repaid itself after seven queries and is pure gain from
the eighth. For an index that serves any
sustained query load this is not a trade-off so much as a rounding error, and
it is the concrete instance of the read/write balance
\cite{valduriez1987} frames in the abstract.

\paragraph{The $N \times M$ column count does not become $N \times M$ storage.}
Materializing one column per segment pair suggests $N \cdot |\Srel| = 6 \times
10^{7}$ slots, six copies of the child collection. It does not cost that,
because a column is $\bot$ wherever the child's parent lives in another parent
segment, and $\bot$ is stored as an absent value rather than a present one:
Lucene's NumericDocValues are sparse (\S\ref{sec:joinindex-impl}). Every child
docid is non-$\bot$ in exactly one of the $N$ columns covering its segment ---
its parent lives in exactly one parent segment --- so the non-$\bot$ entries
over the whole join index number $|\Srel|$, not $N \cdot |\Srel|$. The measured
total, 10{,}000{,}000, is exactly the child collection size, and the auxiliary
index occupies about 30\,MB, i.e.\ some 3 bytes per child document. Storage
grows with the data, not with the segment count; what grows with $N \cdot M$ is
the number of columns to address and sweep, not the bytes they hold.

\paragraph{The build is serial, and need not be.}
All 54 columns were computed one after another on a single thread, in one
batch; no pair was awaited from a concurrent builder. The pairs are
independent, so the 11.3\,s is an artifact of the prototype rather than of the
design --- with the caveat that the write itself must remain a single batch,
since a batch begins at doc $0$ of its sidecar segment, which is what makes a
column's docid coincide with the child docid. Parallelizing the computation is
\S\ref{sec:future}; so is moving the whole cost off the query path into
searcher warming, after which break-even ceases to be a query-latency question
at all.

\subsection{Where the speedup comes from: instrumenting the pruning hierarchy}
\label{sec:pruninginstrumentation}

The latency figure above says the join index is faster; it does not say which
part of the design earns that. To find out, the prototype counts, per parent
segment, how many join-index columns each level of \S\ref{sec:pruning}
eliminates, and how tight the approximation $\Ahat{i}$ actually is.
Table~\ref{tab:pruning-instrumentation} reports those counters over
$3{,}000$ parent-segment scorings of the same benchmark, covering
$27{,}000$ candidate segment pairs.

\begin{table}[H]
\centering
\begin{tabular}{@{}lrr@{}}
\toprule
Measurement & Value & Share \\
\midrule
Candidate pairs $(R_i,S_j)$ with a child-side match & 27{,}000 & \\
\quad dropped a-priori, before any column was opened & 423 & 1.6\% \\
\quad surviving into confirmation                    & 26{,}577 & 98.4\% \\
\midrule
$\Ahat{i}$ as a share of the parent segment (median) & & 100.0\% \\
Range overlap factor $\bigl(\sum_j |[p^{ij}_{\min},p^{ij}_{\max}]|\bigr) / |\Ahat{i}|$ & 8.77 & \\
\midrule
Parent segments where $H_i$ never converged & 86 & 2.9\% \\
Parent segments forced to full convergence  & 2{,}914 & 97.1\% \\
Columns drained                             & 26{,}175 & 98.5\% \\
Columns never opened                        & 402 & 1.5\% \\
Drains ending in an early confirmation      & 3{,}550 & 13.6\% \\
Child docids walked (column-read work)      & 4.34$\times 10^{9}$ & \\
\bottomrule
\end{tabular}
\caption{Pruning effectiveness on the benchmark of
\S\ref{sec:experiments}, over 3{,}000 parent-segment scorings.}
\label{tab:pruning-instrumentation}
\end{table}

Three readings, and they point the same way.

\paragraph{The approximation is vacuous, exactly as predicted.}
The median $\Ahat{i}$ covers $100.0\%$ of its parent segment --- so does the
90th percentile. The approximation never excludes a single parent docid. This
is the prediction of \S\ref{subsec:lazytwophase} confirmed quantitatively:
with one range per column, any column whose parents are spread over $R_i$ has
$p^{ij}_{\min}$ at $0$ and $p^{ij}_{\max}$ at $|R_i|-1$, and the union of nine
such ranges is the whole segment. The overlap factor of $8.77$ says as much
directly: the nine per-column ranges sum to $8.77$ times the size of their
union, i.e.\ each covers essentially all of it.

\paragraph{Document-level pruning therefore collapses.}
Since every non-matching parent that survives the sibling filter is a false
positive, the first one forces convergence: $97.1\%$ of parent segments drained
every column, and only $1.5\%$ of columns went unopened. The $13.6\%$ of drains
that ended in an early confirmation did not save those reads --- they deferred
them, and the segment converged later anyway. The lazy variant of
Algorithm~\ref{alg:lazyhalfread} thus executed, on this workload, as
Algorithm~\ref{alg:base}: $4.34\times10^{9}$ child docids walked is
approximately every matching child docid, once per parent segment.

\paragraph{A-priori pruning fires rarely, for the same reason on the child side.}
Only $1.6\%$ of candidate pairs were bypassed. A column's child range
$[c^{ij}_{\min}, c^{ij}_{\max}]$ is as wide as its parent range for the same
reason, so the bypass has nothing to bite on.

\paragraph{The benchmark is the worst case for range-based pruning, by construction.}
This is not a neutral measurement, and reporting it as one would overstate it in
the other direction. The indexer assigns each sku a foreign key drawn uniformly
at random from all $10^{6}$ products. Every child segment therefore references
parents spread uniformly across every parent segment, which is precisely the
distribution that flattens both $[p^{ij}_{\min}, p^{ij}_{\max}]$ and
$[c^{ij}_{\min}, c^{ij}_{\max}]$ to the full range. Any correlation between
insertion order on the two sides --- a product indexed near its skus, a bulk
load ordered by supplier, anything that puts related documents in nearby
docids --- would tighten both. The numbers above are a lower bound on what the
two range-based levels can do, not an estimate of the typical case, and
quantifying that gap needs a benchmark with a tunable correlation between the
two sides, which \S\ref{sec:futureeval} does not yet include.

\paragraph{What this means for the result.}
The $5.4$--$8.3\times$ speedup of \S\ref{sec:experiments} was obtained with the a-priori
level firing on $1.6\%$ of pairs and the document level saving $1.5\%$ of column
reads. It is therefore not attributable to the pruning hierarchy: it comes from
the join index itself --- the ordinal-to-ordinal column that removes query-time
translation of external keys (R1), and the per-parent-segment execution that
removes the synchronization point of \S\ref{sec:qtimejoin} (R3). That is a
sharper claim than the one we set out to make, and a more robust one: the
structure delivers the speedup on a workload engineered, however
unintentionally, to defeat its optimizations.

It also makes the $n$-range extension (\S\ref{sec:twophase},
\S\ref{sec:apriori}) the highest-value item in \S\ref{sec:future} rather than a
refinement. A single interval per column cannot describe a uniformly scattered
key distribution; a union of disjoint ranges can, and it is the only one of the
proposed changes that attacks the measured cause. Two caveats bound what
Table~\ref{tab:pruning-instrumentation} settles. Segment-level pruning
(\S\ref{sec:segpruning}) is invisible to it, since a segment skipped by the
engine never constructs the context that would report it; and the $50.6\%$ of
confirmations answered from $H_i$ alone is dominated by lookups made after
convergence, when $H_i = P_i$ and the question is no longer being pruned but
merely answered.

\subsection{Future Evaluation Plan}
\label{sec:futureeval}

\begin{itemize}
    \item vary the correlation between the two sides' insertion orders. The
          current generator assigns foreign keys uniformly at random, which
          flattens every column's parent and child ranges and reduces both
          range-based pruning levels to noise
          (\S\ref{sec:pruninginstrumentation}); a tunable locality parameter
          would measure them across the range from that worst case to the
          co-located case, and would tell us what the $n$-range extension has
          to beat;
    \item benchmark under a moderate stream of concurrent updates;
    \item add a \emph{global-ordinals baseline arm} --- Lucene's global-ordinals
          join (\S\ref{subsec:otherjoins}), reachable in Apache Solr as
          \texttt{method=topLevelDV}. This is the comparison the paper most
          needs and does not have. Global ordinals attack the same cost that
          R1 (\S\ref{subsec:requirementsandresults}) identifies in the classic
          query-time join --- they correlate the two sides' value ordinals once,
          up front, instead of translating external keys per query --- so they
          are the strongest same-system baseline for the central claim, and
          measuring against \texttt{\{!join score=none\}} alone leaves that
          claim weaker than it needs to be. We expect the comparison to split by
          workload rather than to settle. A global ordinal map is built per
          top-level reader and invalidated on every commit, so its advantage
          should be largest on a static index and should erode as the update
          rate rises, while a join-index column is scoped to an immutable
          segment \emph{pair} and survives commits that do not delete either
          segment. The two arms should therefore be swept against the update
          stream of the item above rather than measured at a single, static
          operating point --- and the same sweep bounds what the global-ordinals
          restriction to one index and one field
          (\S\ref{subsec:otherjoins}) costs in exchange;
    \item compare with memory buffer directory and soft-commits (NRT-search);
    \item compare with joins over numeric fields;
    \item compare against a state-of-the-art RDBMS join.
\end{itemize}

\section{Discussion}
\label{sec:discussion}

This approach might be applicable to other LSM-tree storage
engines and/or those which are based on virtual IDs.

\section{Future Work}
\label{sec:future}

We group the outstanding work by how much of it is already decided, rather
than by topic: what remains to be built on the current prototype, what is a
charted extension of it, and what is still an open question.

\paragraph{Engineering on the current prototype.}
Bounded work whose shape is already clear from \S\ref{sec:implementation}:
\begin{itemize}
    \item cache the column-to-auxiliary-index-segment assignment (so far it searches columns by scanning segments linearly);
    \item offload the join-index build onto searcher warmup, which should remove the tail latency observed in \S\ref{sec:experiments};
    \item explore parallelism when writing join-index columns. The bulk pass of
          \S\ref{sec:joinindex-impl} computes each pair's column on the calling thread, one pair
          after another, although the pairs are independent: each merges the term dictionaries of
          its own $(R_i, S_j)$ and touches no shared state but the scratch buffer, which can be
          made per-thread. The constraint is that the write itself must stay a single batch --- a
          batch begins at doc $0$ of its sidecar segment, which is what makes a column's docid
          coincide with the child docid --- so the parallelism belongs in computing the mappings,
          not in writing them;
    \item add join-index column cardinality to the metadata alongside the bounding ranges, and use it to order child segments in lazy confirmation (\S\ref{subsec:lazytwophase});
    \item improve the sweeping and merging scheme for auxiliary index columns.
\end{itemize}

\paragraph{Charted extensions.}
Larger, but the direction is settled:
\begin{itemize}
    \item extend bounding ranges to a union of $n$ disjoint ranges, to tighten the parent docid approximation and allow leapfrogging of child matches (\S\ref{sec:twophase}, \S\ref{sec:apriori}). The measurements of \S\ref{sec:pruninginstrumentation} make this the first item to attempt rather than one among several: it is the only proposed change that addresses why both range-based levels currently fire so rarely;
    \item extend the current 1:M onto a generic N:M relation;
    \item semijoin in the reverse direction --- return children of certain parents via the same join-index columns;
    \item prototype a real join operation returning the children side as well;
    \item apply the join index to generic graph traversal\footnote{\url{https://solr.apache.org/guide/solr/latest/query-guide/other-parsers.html\#graph-query-parser}};
    \item introduce joins in the existing SQL parser\footnote{\url{https://solr.apache.org/guide/solr/latest/query-guide/sql-query.html}}.
\end{itemize}

\paragraph{Open questions.}
Where the aim itself is still to be established:
\begin{itemize}
    \item complexity analysis, and with it a read/write cost model in Valduriez's sense~\cite{valduriez1987} --- when materializing $N \times M$ columns pays for itself against joining at query time;
    \item following from that, whether the choice can be made adaptively, falling back to the classic query-time join (\S\ref{sec:qtimejoin}) as the segment count $N$ grows and the redundant child-query evaluations of \S\ref{sec:base} come to dominate;
    \item explore indexing multiple relation joins;
    \item research compatibility with vector search (HNSW\footnote{\url{https://solr.apache.org/guide/solr/latest/query-guide/dense-vector-search.html\#query-time}});
    \item introduce specialized data structures for a faster join;
    \item research a file format allowing the join index to be kept on one side, without an auxiliary index.
\end{itemize}

\section{Conclusion}
\label{sec:conclusion}

We have shown how Valduriez's join-index idea, originally developed for relational
systems, carries over to the flush-produced, unmerged segments of an inverted
index, using Apache Lucene as a concrete implementation target. On top of this
structure we built a semijoin algorithm that runs per parent segment in
parallel, without a global barrier between stages, and prunes at three levels
--- segment-level, which comes free from per-segment execution; a-priori
min/max; and document-level two-phase confirmation
--- so that it composes with the dynamic pruning already present in search
engines rather than fighting it. On a benchmark of 1M products and 10M skus
(\S\ref{sec:experiments}), this join index cut average query latency
$5.4\times$ (359.8\,ms vs.\ 1934.6\,ms) against Solr's built-in query-time
join, and $8.3\times$ once eight queries run at once --- a point at which the
baseline has passed its own throughput peak and begun to decline, while the
join index is still gaining --- all while returning an identical result count
on all 500 queries of every run.

Instrumenting that run (\S\ref{sec:pruninginstrumentation}) places the credit
more precisely than we first expected. On a workload whose foreign keys are
uniformly random --- and therefore flatten every column's bounding ranges to
the full segment --- the a-priori level bypassed $1.6\%$ of segment pairs and
the document level spared $1.5\%$ of column reads, so the speedup is the join
index's own, not its pruning's. We regard that as the stronger outcome: the
ordinal-to-ordinal column and the synchronization-free per-segment execution
carry the result unaided, and the pruning hierarchy remains available for the
correlated key distributions where its single-interval approximation has
something to exclude.

\appendix

\section{Appendix}
\label{app:notes}

\subsection{Inverted Index as Dictionary Compression}
\label{app:invidx_dict}
If we precisely look into the first point of \S\ref{sec:valueordinals}, we can recognize
the dictionary compression~\cite{abadi2013columnstores} scheme in inverted index.

To keep the argument in its simplest form, we restrict it to a
\emph{primary-key} field --- exactly the case of $\R.id$ in this paper --- and
assume throughout this subsection that
\begin{itemize}
    \item $f$ is single-valued: every docid has exactly one value ordinal;
    \item values are unique: every ordinal occurs in exactly one document,
          so a posting list is a single docid;
    \item the term dictionary is a collision-free in-memory hash table, so
          the ordinal of a value $v$ is
          $\texttt{ords}^{R_i}_{f}[v] = \texttt{ordinals}[\,\mathit{hash}(v)
          \bmod \mathit{tableSize}\,]$. For sure, real implementations are much more complex than that.
\end{itemize}
Under these assumptions, keeping the notation of
\S\ref{subsec:columnsandquery}, the two access paths are:
\begin{itemize}
    \item searching, i.e.\ obtaining a docid by a value, is
          $\texttt{docids}^{R_i}_{f}\bigl[\texttt{ords}^{R_i}_{f}[v]\bigr]$;
    \item getting a value by docid is
          $\texttt{values}^{R_i}_{f}\bigl[R_i.f[p]\bigr]$.
\end{itemize}
The two are not merely symmetric, they are mutually inverse bijections
between docids and ordinals:
\[
\texttt{docids}^{R_i}_{f}\bigl[R_i.f[p]\bigr] = p
\qquad\text{and}\qquad
R_i.f\bigl[\texttt{docids}^{R_i}_{f}[o]\bigr] = o .
\]
In words: the inverted index is the column read backwards. The column
$R_i.f$ maps docids to ordinals, the inverted index maps ordinals back to
docids, and $\texttt{values}^{R_i}_{f}$, $\texttt{ords}^{R_i}_{f}$ are
likewise inverse to each other between ordinals and values. So the inverted
index and the dictionary-compressed column are two indexings of one and the
same correspondence between docids and values: the column stores it indexed
by docid, the inverted index stores it indexed by value.

Dropping the assumptions does not change the picture, only weakens the
one-to-one correspondence into a many-to-one one: if values repeat, the inverted index becomes the
set-valued inverse of the column --- the posting list of an ordinal is the
\emph{preimage} $(R_i.f)^{-1}(o)$, and posting-list length is exactly the
redundancy that dictionary compression exploits; if the field is
multi-valued, the column becomes set-valued in the same way.

The following observations demonstrate tight coupling of inverted index and data columns, which are formally distinct entities in Lucene right now:
\begin{itemize}
    \item long ago Lucene obtained the data column at runtime via \emph{uninverting}
    \footnote{\url{https://lucene.apache.org/core/6_3_0/misc/org/apache/lucene/uninverting/UninvertingReader.html}}
    meaning the process of reversing inversion (of inverted index). Note: nowadays it's replaced by
    index-time data columns called by DocValues, where SortedDocValues
    introduce dictionary compression explicitly;
    \item there's a proposal\footnote{\url{https://issues.apache.org/jira/browse/LUCENE-5832} - Explore a combined DocValues/PostingsFormat that shares a single terms dict} 
    about merging inverted index's term dictionary and data column's dictionary into a single one;
    \item pulsing codec~\cite{cutting1990dynamic} is exact mapping of value to docid;
    \item possible contra argument about a posting list arity can be debated with examples
    of multi-valued data columns (SortedSetDocValues\footnote{\url{https://lucene.apache.org/core/10_1_0/core/org/apache/lucene/document/SortedSetDocValuesField.html}}) and from other side,
    the pulsing codec emphasizes the importance of the singular case.
\end{itemize}

Overall, this observation does not bring much to practice, but it may give us a clue in designing
new data structures and indices formats.

\clearpage
\section{Symbol Table}
\label{app:symbols}

\begin{table}[H]
\centering
\caption{Notation used throughout the paper.}
\label{tab:symbols}
\begin{tabular}{ll}
    \toprule
    \textbf{Symbol} & \textbf{Reads as} \\
    \midrule
    $\R$ & parent (outer side) relation, "to" side  \\
    $\R.id$ & parent side primary key  \\
    $\Srel$ & child, inner relation, "from" side \\
    $\Srel.fk$ & child side foreign key \\
    $q_R$, $q_S$ & parent / child query \\
    $\ltimes$ & semijoin \\
    $R_i$, $S_j$; $N$, $M$ & segments and their counts \\
    $p$, $c$ & parent / child docids \\
     & (document ordinal numbers) \\
    $C$, $U_j$ & a set of child docids; the docid space $[0,|S_j|)$ \\
    $R_i.f[p]$ & positional column access \\
     & $f$ field value ordinal by $p$ docid \\
    $\texttt{values}^{R_i}_{f}[\texttt{v\_ordinal}]$ & obtaining $f$ field value by ordinal \\
    $\texttt{ords}^{R_i}_{f}[v]$ & term-dictionary lookup: ordinal by value \\
    $\texttt{docids}^{R_i}_{f}[o]$ & posting list: docids by ordinal \\
     & (\S\ref{app:invidx_dict}, inverse of $R_i.f$) \\
    $\M(q, S_j)$ & sorted match set / iterator \\
    $\mathit{next}/\mathit{advance}$ & iterator contract \\
    $\mathit{approximation}/\mathit{matches}$ & (approx./confirm.\ phases) \\
    $\Ji$, $\bot$ & join-index column; absent-parent sentinel \\
    $p^{ij}_{\min}, p^{ij}_{\max}, c^{ij}_{\min}, c^{ij}_{\max}$ & column parent/child, min/max metadata \\
    $P_i$, $\Ahat{i}$, $H_i$ & exact / approximate / half-read set \\
    \bottomrule
\end{tabular}
\end{table}

\section{Worked Example}
\label{app:example}

We trace the algorithms of \S\ref{sec:algorithm} end-to-end on a
tiny instance with $N{=}2$ parent segments and $M{=}3$ child
segments. The instance is small but not degenerate: it exercises the
a-priori level of \S\ref{sec:apriori} in all three of its uses
(\S\ref{app:example-base}), and it contains a false positive that puts the
lazy variant into both of its regimes --- early confirmation for $R_1$,
forced convergence for $R_2$ (\S\ref{app:example-lazy}).

\subsection{Setup}
\label{app:example-setup}

Parent segment $R_1$ holds three docs, local docids
$p=0,1,2$, with ids $A,B,C$; parent segment $R_2$ holds three docs,
local docids $p=0,1,2$, with ids $D,E,F$. Three child segments
$S_1, S_2, S_3$ hold, respectively, $4, 3, 4$ docs, each carrying a
foreign key pointing at one of $A..F$:
\[
\begin{array}{c|cccc}
S_1,\ c: & 0 & 1 & 2 & 3 \\
\mathit{fk}: & B & D & A & F
\end{array}
\qquad
\begin{array}{c|ccc}
S_2,\ c: & 0 & 1 & 2 \\
\mathit{fk}: & E & A & C
\end{array}
\qquad
\begin{array}{c|cccc}
S_3,\ c: & 0 & 1 & 2 & 3 \\
\mathit{fk}: & F & B & D & E
\end{array}
\]
A single child query $q_S$ matches
$\M(q_S,S_1) = \{0,2,3\}$, $\M(q_S,S_2) = \{1,2\}$,
$\M(q_S,S_3) = \{0,1,3\}$ (sorted iterators, in this order).

\subsection{Join-index columns}
\label{app:example-columns}

By Definition~\ref{def:joinindexcolumn}, each of the $N{\times}M = 6$ columns $J_{ij}$
maps a child docid of $S_j$ to a local parent docid of $R_i$, or
$\bot$ if the fk target does not live in $R_i$:
\[
J_{11} = (1,\bot,0,\bot), \quad
J_{12} = (\bot,0,2), \quad
J_{13} = (\bot,1,\bot,\bot),
\]
\[
J_{21} = (\bot,0,\bot,2), \quad
J_{22} = (1,\bot,\bot), \quad
J_{23} = (2,\bot,0,1).
\]
The associated metadata (over non-$\bot$ entries only) is
\begin{center}
\begin{tabular}{lcccc}
\toprule
column & $c_{\min}$ & $c_{\max}$ & $p_{\min}$ & $p_{\max}$ \\
\midrule
$J_{11}$ & 0 & 2 & 0 & 1 \\
$J_{12}$ & 1 & 2 & 0 & 2 \\
$J_{13}$ & 1 & 1 & 1 & 1 \\
$J_{21}$ & 1 & 3 & 0 & 2 \\
$J_{22}$ & 0 & 0 & 1 & 1 \\
$J_{23}$ & 0 & 3 & 0 & 2 \\
\bottomrule
\end{tabular}
\end{center}

\subsection{Approximation phase}
\label{app:example-approx}

Unioning the per-column $[p_{\min},p_{\max}]$ ranges over $j=1..3$
(Algorithm~\ref{alg:twophase_alg}, line~1) gives
\[
\Ahat{1} = [0,1] \cup [0,2] \cup [1,1] = \{0,1,2\},
\qquad
\Ahat{2} = [0,2] \cup [1,1] \cup [0,2] = \{0,1,2\}.
\]
Both approximations cover all of their segment ($R_1$ and $R_2$
have only 3 docs each), so no candidate is pruned yet by
\S\ref{sec:apriori} at this granularity --- but $\Ahat{2}$ already
contains a false positive, $p{=}0$ ($D$), that the confirm phase
below must rule out.

\subsection{Ground truth via the base algorithm}
\label{app:example-base}

For reference, Algorithm~\ref{alg:base} gives the exact semijoin
sets directly (writing only the non-$\bot$ entries that each column
contributes):
\[
P_1 = \{J_{11}[0], J_{11}[2]\} \cup \{J_{12}[1], J_{12}[2]\}
      \cup \{J_{13}[1]\}
    = \{1,0\} \cup \{0,2\} \cup \{1\} = \{0,1,2\},
\]
\[
P_2 = \{J_{21}[3]\} \cup \{\} \cup \{J_{23}[0], J_{23}[3]\}
    = \{2\} \cup \{2,1\} = \{1,2\}.
\]
So every doc of $R_1$ matches ($A,B,C$), while $R_2$ matches only
$E,F$ --- $D$ is a match for no child in $M(q_S,\cdot)$, despite
sitting inside $\Ahat{2}$.

The a-priori guards of \S\ref{sec:apriori} are already in force here, and
each of their three uses fires at least once. The
$\mathit{advance}(c^{ij}_{\min})$ entry skips a child match before any read:
$J_{21}$ is entered at $c{=}1$ and never sees the match $c{=}0$, as is
$J_{13}$. The $c > c^{ij}_{\max}$ exit abandons a column early: $J_{11}$
stops at $c{=}3$ and $J_{13}$, defined only at $c \in [1,1]$, contributes
its single entry and retires, reading one of the three matches of $S_3$.
Finally, the empty term of $P_2$ is the bypass: $J_{22}$ is defined only at
$c{=}0$, which $q_S$ does not match, so $\mathit{advance}(0)$ returns
$c{=}1 > c^{22}_{\max}{=}0$ and the column is never opened --- that empty
set costs one $\mathit{advance}$ and no column read.

\subsection{Lazy confirmation trace (Algorithm~\ref{alg:lazyhalfread})}
\label{app:example-lazy}

Each parent segment keeps its own $H_i$ and its own persisted
iterators $it_j$ over $\M(q_S,S_j)$, $j=1,2,3$;
$\mathit{matches}(p)$ is invoked once for each local docid of $R_i$
inside $\Ahat{i}$, in ascending order. We apply the a-priori guards of
Algorithm~\ref{alg:lazyhalfread}: the first $\mathit{step}$ on a child segment is
$\mathit{advance}(c^{ij}_{\min})$, and the inner loop retires the segment
once $c$ passes $c^{ij}_{\max}$, using the metadata tabulated in
\S\ref{app:example-columns}.

\paragraph{$R_1$: early breaks pay off.} Start with $H_1=\emptyset$,
$\mathit{unread}=[1,2,3]$.
\begin{itemize}
\item $\mathit{matches}(0)$: $0 \notin H_1$. Read $it_1$, entered at
      $\mathit{advance}(c^{11}_{\min}{=}0)$: $c{=}0
      \Rightarrow J_{11}[0]{=}1$, $H_1 \leftarrow \{1\}$ (no match
      yet); $c{=}2 \Rightarrow J_{11}[2]{=}0$, $H_1 \leftarrow
      \{0,1\}$ --- $0{=}0$, \textbf{return \textsc{true}}. $it_1$ is left
      paused at $c{=}2$ consumed, $c{=}3$ still unread; $S_2,S_3$
      untouched.
\item $\mathit{matches}(1)$: $1 \in H_1$ already --- \textbf{return
      \textsc{true}} for free, no column read at all.
\item $\mathit{matches}(2)$: $2 \notin H_1=\{0,1\}$. Resume $it_1$
      at $c{=}3$, which exceeds $c^{11}_{\max}{=}2$: $S_1$ retires
      without $J_{11}[3]$ ever being read, pop,
      $\mathit{unread}=[2,3]$. Move to $j{=}2$, entered at
      $\mathit{advance}(c^{12}_{\min}{=}1)$:
      $c{=}1 \Rightarrow J_{12}[1]{=}0$ (already in $H_1$, no
      match); $c{=}2 \Rightarrow J_{12}[2]{=}2$, $H_1 \leftarrow
      \{0,1,2\}$ --- $2{=}2$, \textbf{return \textsc{true}}. $J_{13}$
      ($S_3$) is never read.
\end{itemize}
$H_1$ converges to $\{0,1,2\} = P_1$, matching \S\ref{app:example-base},
after touching only 4 of the $3{+}2{+}3{=}8$ available child
matches --- three saved by the early returns, one by the
$c^{11}_{\max}$ bound.

\paragraph{$R_2$: a false positive forces a full drain.} Start with
$H_2=\emptyset$, $\mathit{unread}=[1,2,3]$.
\begin{itemize}
\item $\mathit{matches}(0)$ (the false positive $D$ from
      $\Ahat{2}$): $0 \notin H_2$, and no read of any column can
      ever confirm it, so the loop must retire \emph{all} three
      child segments before giving up. $j{=}1$, entered at
      $\mathit{advance}(c^{21}_{\min}{=}1)$, which skips the match
      $c{=}0$ outright: $c{=}2 \Rightarrow J_{21}[2]{=}\bot$;
      $c{=}3 \Rightarrow J_{21}[3]{=}2$, $H_2 \leftarrow \{2\}$
      ($2 \neq 0$); $it_1$ exhausted, pop. $j{=}2$:
      $\mathit{advance}(c^{22}_{\min}{=}0)$ yields $c{=}1$, already
      past $c^{22}_{\max}{=}0$, so the column is bypassed and $J_{22}$
      is never loaded at all --- though $S_2$'s child matches are
      materialized regardless (\S\ref{sec:base}); pop. $j{=}3$, entered at
      $\mathit{advance}(c^{23}_{\min}{=}0)$: $c{=}0
      \Rightarrow J_{23}[0]{=}2$ ($H_2$ unchanged); $c{=}1
      \Rightarrow J_{23}[1]{=}\bot$; $c{=}3 \Rightarrow
      J_{23}[3]{=}1$, $H_2 \leftarrow \{1,2\}$ ($1 \neq 0$); $it_3$
      exhausted, pop. $\mathit{unread}=\emptyset$ ---
      \textbf{return \textsc{false}}.
\item $\mathit{matches}(1)$: $1 \in H_2=\{1,2\}$ --- \textbf{return
      \textsc{true}} for free.
\item $\mathit{matches}(2)$: $2 \in H_2$ --- \textbf{return \textsc{true}}
      for free.
\end{itemize}
Here the single false positive forces convergence up front (the worst case
of \S\ref{subsec:lazytwophase}), and the a-priori level is all that stands
between it and a full read: of the eight child matches it costs five column
lookups, and $J_{22}$ --- whose single defined entry lies outside the child
query's matches --- is bypassed without being opened. That cost is paid
exactly once per segment: $H_2$ has now fully converged to
$P_2 = \{1,2\}$, and the two remaining confirmations are free lookups.

\bibliographystyle{plain}
\bibliography{references}

\end{document}